\documentclass[aps,twocolumn]{revtex4}

\usepackage{graphicx}
\usepackage{dcolumn}
\usepackage{bm}
\usepackage{amsmath}
\usepackage{amssymb}

\def \ba {\begin{eqnarray}}
\def \ea {\end{eqnarray}}

\begin{document}



\title{The maximal superconductivity in proximity to charge density wave quantum critical point in Cu$_x$TiSe$_2$ }

\author{Tae-Ho Park}
\email{thpark@skku.edu}
\affiliation{
 Department of Physics, Sungkyunkwan University, Suwon 16419, Korea. }

\author{Obinna P. Uzoh}
\affiliation{
 Department of Physics, Sungkyunkwan University, Suwon 16419, Korea. }

\author{Han-Yong Choi}
\email{hychoi@skku.edu}
\affiliation{
 Department of Physics, Sungkyunkwan University, Suwon 16419, Korea. \\
 Asia Pacific Center for Theoretical Physics, Pohang 37673, Korea.}

\date{\today}

\begin{abstract}

Superconductivity emerges in $1T$-TiSe$_2$ when its charge density wave (CDW) order is suppressed by Cu intercalation or pressure.
Since the CDW state is thought to be an excitonic insulator, an interesting question is whether the superconductivity is also mediated by the excitonic fluctuations.
We investigated this question as to the nature of doping induced superconductivity in Cu$_x$TiSe$_2$ by asking if it is consistent with the phonon-mediated pairing.
We employed the {\it ab initio} density functional theory and density functional perturbation theory to compute the electron-phonon coupling Eliashberg function from which to calculate the superconducting (SC) critical temperature $T_c$.
The calculated $T_c $ as a function of the doping concentration $x$ exhibits a dome shape with the maximum $T_c$ of $2-6$ K at $x \approx 0.05$ for the Coulomb pseudopotential $0 \leq \mu^* \leq 0.1$. The maximal $T_c$ was found to be pinned to the quantum critical point at which the CDW is completely suppressed and the corresponding phonon mode becomes soft. Underlying physics is that the reduced phonon frequency enhances the electron-phonon coupling constant $\lambda$ which overcompensates the frequency decrease to produce a net increase of $T_c$.
The doping induced superconductivity in Cu$_x$TiSe$_2$ seems to be consistent with the phonon-mediated pairing. Comparative discussion was made with the pressure induced superconductivity in TiSe$_2$.

\end{abstract}

\maketitle

\section{Introduction}

The interplay between superconductivity and other orders in proximity such as antiferromagnetism, nematicity, and CDW has been one of engaging research topics in condensed matter physics.\cite{Testardi_RMP1975,Kiss_Naturephys2007,Loret_2019,jiang2021,Otto_Scienceadv2021} Vast classes of superconductors including the copper oxides, Fe based superconductors, heavy fermion materials, and transition metal dichalcogenides (TMDC), despite their diversity in crystal structure and phenomenology, exhibit the common trait that the SC critical temperature $T_c$ shows a dome shape as the tuning parameters like the chemical concentration, pressure, and external field are varied. The maximum $T_c$ appears close to a quantum critical point (QCP) at which one of the neighboring orders is completely suppressed as a tuning parameter is varied. This is reminiscent of the quantum critical superconductivity which posits that the very interaction underlying the neighboring order also induces the superconductivity. The recurrence of this ``universal phase diagram'' suggests a deep connection between superconductivity and neighboring orders.

\begin{figure}
 \includegraphics[width=\linewidth]{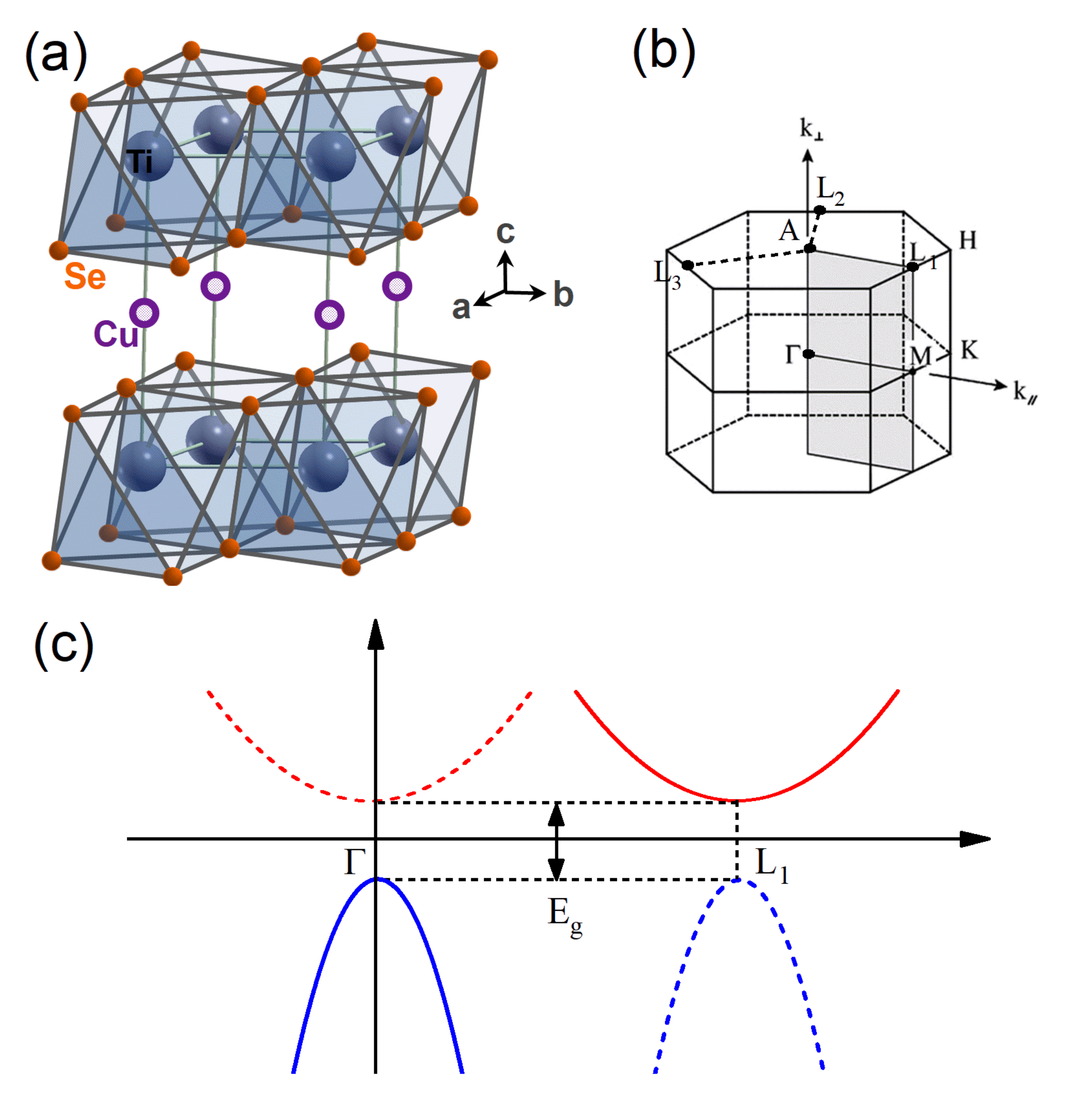}
\caption{(a) The crystal structure of Cu$_x$TiSe$_2$ which illustrates the intercalated Cu atoms in between the TiSe$_2$ layers. (b) The 1st BZ of $1T$-TiSe$_2$ corresponding to the crystal structure of (a). (c) The energy dispersion of the hole and electron bands along the $\Gamma-L_1$ direction. The hole band around the $\Gamma$ and three electron bands around $L_i$ points are shown in blue and red, respectively. $E_g$ is the indirect gap between the electron and hole bands. The BZ is reduced in the CDW phase and appear the backfolded dispersions as shown in the blue and red dashed lines.}
\label{fermiband}
\end{figure}

The TMDC materials MX$_2$, where M = Nb, Ti, Ta, Mo, and X = Se, S, exhibit this phase diagram out of interplay between superconductivity and CDW order. Of particular interest is $1T$-TiSe$_2$ because its CDW is thought to be induced by the exciton condensation of electron-hole pairs.
The $1T$-TiSe$_2$ is a TMDC semiconductor/semimetal of a layered structure with an indirect gap/overlap between the Se $4p$ hole band centered at the $\Gamma$ point and the Ti $3d$ electron bands around the $L$ points in the Brillouin zone (BZ) as shown in Fig.\ \ref{fermiband}.
An exciton, a bound state of an electron from the $L_i$ band and a hole from $\Gamma$ band in Fig.\ \ref{fermiband} (b), then has a nonzero net momentum and the inverse of the momentum sets a new length scale. Consequently, the exciton condensation is accompanied by a structural instability at the inverse momentum and makes a phase transition to a CDW of $2\times2\times2$ superstructurebelow the critical temperature $T^{CDW} \approx 200$ K.\cite{DiSalvo1976}

The CDW is suppressed by Cu intercalation or pressure or electric field and emerges superconductivity. The SC critical temperature $T_c$ exhibits a dome shape as a function of the Cu concentration $x$ or the pressure $P$. As for the Cu intercalation, $T_c$ becomes maximum of $3.79-4.15$ K at $x_{opt} \approx 0.077-0.08$ in close proximity to the QCP of $x_c \approx 0.06-0.07$ at which the CDW is completely suppressed.\cite{EMorosan06,Sawa19}
Also, Raman scattering experiments showed the frequency softening and divergent linewidth of the CDW amplitude mode
corresponding to the $L$ point as $x$ approaches $x_c$.\cite{Barath2008,Holy_PRB1977}
These seem in accord with the quantum critical superconductivity alluded above which suggests that the very interaction underlying the CDW formation may also induce the superconductivity in Cu$_x$TiSe$_2$.
An alternative view is that the proximity of CDW and SC is coincidental and the superconductivity is phonon-mediated conventional one.\cite{Zhao_PRL2007}

As to the pressure, Kusmartseva $et~al.$ observed by transport measurements that superconductivity appears in the range of $P \approx 2-4$ GPa and the maximum $T_c \approx 1.8$ K occurs around $P_{opt} \approx 3$ GPa close to the CDW suppression.\cite{Kusmartseva2009} However, Joe {\it et al.} \cite{YIJoe14} and Kitou {et al.} \cite{Kitou_PRB2019} observed with the synchrotron X-ray diffraction on single crystals
that the $P$ induced suppression is at $P_c \approx 5.1 $ GPa which is more than 1 GPa beyond the end of the SC region.
Joe {\it et al.} also observed a reentrant incommensurate CDW phase appeared near the $P_{opt} $ above the SC dome which seemed to indicate that the pressure induced superconductivity in TiSe$_2$ may not be connected to the CDW suppression but to the CDW domain walls. Calandra and Mauri showed with the $ab~initio$ calculations that the behavior of $T_c$ as a function of pressure is entirely determined by the electron-phonon interaction without a need for invoking exciton mechanism.\cite{Calandra2011}

For intercalated Cu$_x$TiSe$_2$, Zhao {\it et al.}\cite{Zhao_PRL2007} and Qian {\it et al.}\cite{Qian2007} observed with ARPES experiments that the Cu doping raises the chemical potential from inside the semiconducting gap at zero doping to the Ti $3d$ electron band. This of course leads to the enhanced electronic density of states (DOS) and the increased screening, which enhance superconductivity and weaken exciton condensation, respectively. These are the two different effects of the doping by which the authors argued that the seeming competition between CDW and SC is coincidental.
Li {\it et al.} measured the in-plane thermal conductivity as a function of temperature for Cu$_x$TiSe$_2$ at $x=0.06$ and concluded that it is a conventional $s$-wave single gap superconductor.\cite{Li2007}
Kogar {\it et al.} performed X-ray diffraction experiments on Cu$_x$TiSe$_2$ and proposed that the incommensuration and domain walls of the CDW may play a crucial role in the formation of the SC state as in the pressure induced SC mentioned above.\cite{Kogar_PRL2017}

On the other hand, Kitou {\it et al.}\cite{Kitou_PRB2019} and Maschek {\it et al.}\cite{Maschek_PRB2016} stressed the dissimilarity between the doping induced and pressure induced cases. For instance, Maschek {\it et al.} argued that the doping induced SC can be understood by the phonon mediated pairing mechanism alone while a hybridization of phonon and exciton modes is necessary for the pressure induced SC.
The possibility of an unconventional $s_{\pm}$ pairing between the electron and incipient hole bands and the time-reversal-symmetry breaking chiral superconductivity has also been proposed out of the interplay between CDW and pairing.\cite{Ganesh_PRL2014} See below.


Here, we will demonstrate that the superconducting $T_c$ as a function of the Cu intercalation concentration $x$ for Cu$_x$TiSe$_2$ is determined by the electron-phonon interaction based on the {\it ab initio} density functional theory (DFT) and density functional perturbation theory (DFPT) calculations which consider the Cu intercalation explicitly.
From these calculations the maximal $T_c$ was found to be pinned to the minimum phonon frequency. This naturally leads to the dome shape of $T_c$ as a function of an external parameter as has been observed and discussed in many classes of materials.\cite{Testardi_RMP1975,Moussa_PRB2006} Underlying physics will be discussed after the presentation of the calculations.
We argue that the superconductivity in Cu$_x$TiSe$_2$ is phonon mediated conventional pairing and the proximity of the maximal $T_c$ and the CDW quantum critical point is not coincidental.

\section{Computational Details}

\begin{figure}[t]
\includegraphics[width=1\linewidth]{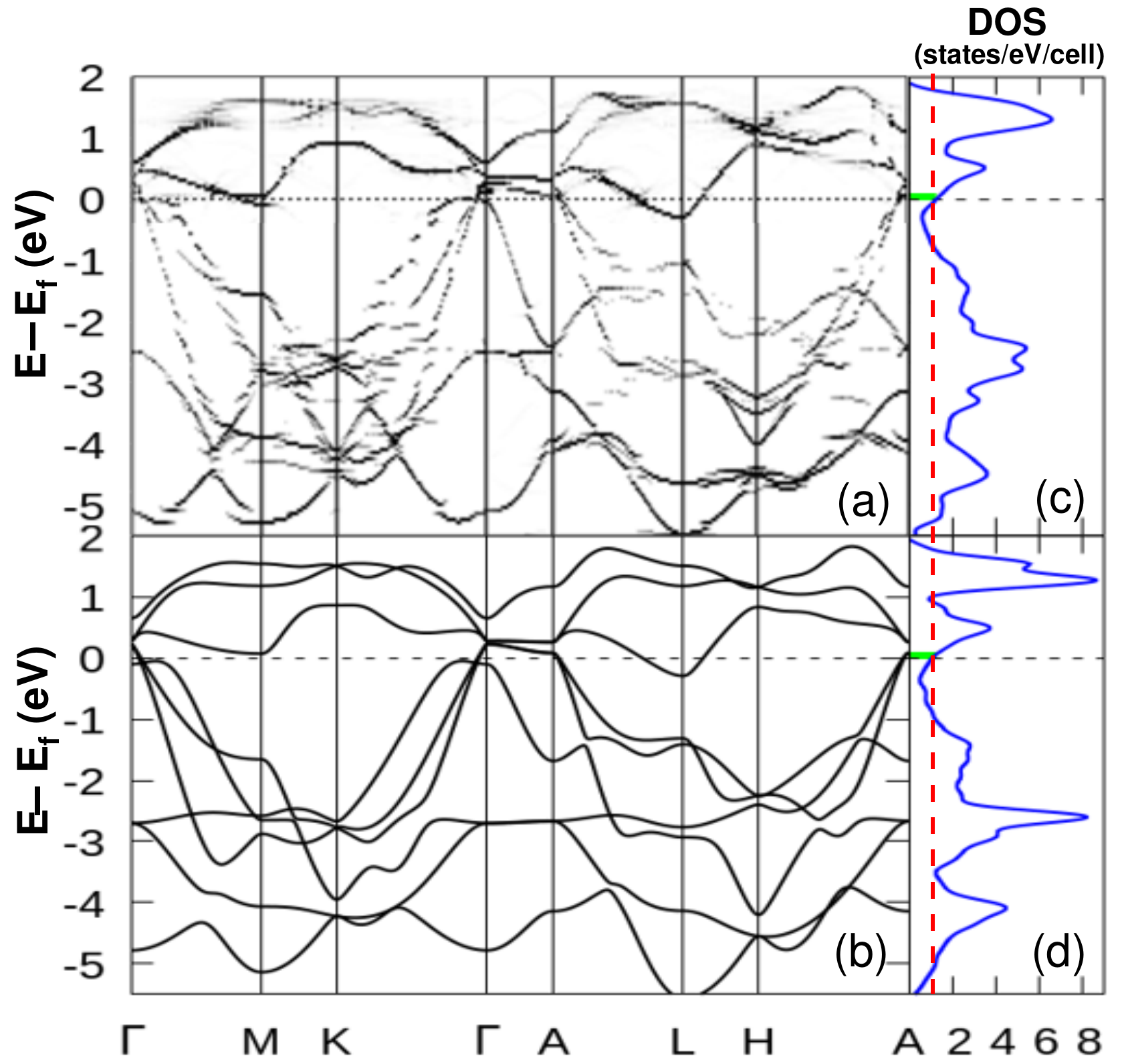}
\caption{\label{figure1}
(a) The electronic band structures of unfolded $3\times3\times1$ supercell $\mathrm{{Cu}}_{0.11}\mathrm{{TiSe}}_{2}$, and (b) VCA $\mathrm{{Li}}_{0.11}\mathrm{{TiSe}}_{2}$.
(c) The electronic DOS for the supercell system with the green area around the Fermi level corresponding to the amount of electrons transferred from the Cu cations to the $\mathrm{{TiSe}}_{2}$ layers, and
(d) DOS for VCA system with the green area corresponding to the amount of electrons transferred from the virtual Li cations to the $\mathrm{{TiSe}}_{2}$ layers. The vertical (red dashed) line is a guide to eyes to compare
DOS at the Fermi level for supercell to VCA calculations.
}
\label{fig2_DOS}
\end{figure}

We performed all electronic and phonon computations using first-principles methods within the
QUANTUM ESPRESSO suite,\cite{Giannozzi2009} implementing the DFT. For the exchange-correlation contribution to the total energy, we used the local density approximation (LDA) functional in the parameterization of Perdew-Zunger.\cite{JPPerdew1981}
For all the systems calculated, the lattice parameters were fixed to the
experimentally observed values obtained in Ref. \cite{EMorosan06} and the atomic
positions were then allowed to fully relax until the forces on the atoms became
less than $10^{-5}$ Ry/a.u.
In the electronic structure calculations, core electrons were treated
with the optimized norm-conserving Vanderbilt pseudopotentials, \cite{Hamann_PRB2013}
while the Ti $3s^2 3p^6 4s^2 3d^2$ and
Se $3d^{10} 4s^2 4p^4$ were treated as valence electrons and was described
with plane waves up to a kinetic energy cutoff of 70 Ry. The BZ of the
primitive cell was sampled with a $24\times24\times24$ Monkhorst-Pack
\cite{HJMonkhorst1976} \textbf{k}-point grid using a Marzari-Vanderbilt \cite{Marzari99}
smearing of 0.01 Ry.

To take the doping effects into consideration for Cu$_x$TiSe$_2$, we utilized the virtual crystal approximation (VCA) technique.\cite{Ramer&Rappe2000,Bellaiche2000} In the VCA, the system under study is computed in the primitive periodicity of the crystal with a virtual atom that interpolates between two constituent atoms. It is computationally much less expensive in comparison with, for example, the supercell approach which requires very large supercells.\cite{Jishi2008} Within a pseudopotential approach to DFT, the pseudopotential $V_{ps}^{VA}$ for a virtual atom can be constructed by a simple superposition of the pseudopotentials $V_{ps}^{A}$ and $V_{ps}^{B}$ of two atoms $A$ and $B$ as\cite{Ramer&Rappe2000,Bellaiche2000}
\begin{equation}
 V_{ps}^{VA} = xV_{ps}^A + (1 - x)V_{ps}^B.
\label{eq:1VCA}
\end{equation}
However, using Cu atom within VCA to reproduce the electronic structure of
more accurate supercell calculations is problematic, as was also reported in Ref.
\cite{SawaRapid2019}.
Cu outer $4s$ orbital participates in the electron doping of TMDC materials,\cite{Jishi2008, Muhammad2018}
and the Cu $3d$ electrons are irrelevant in the electronic structure near
the Fermi level. Sawa {\it et al.} demonstrated that virtual Li atom can successfully simulate the Cu doping because Li has one electron
in the $s$ outer shell like Cu within the VCA technique.\cite{SawaRapid2019} Following this, we constructed virtual pseudopotentials by mixing Li and He norm-conserving pseudopotentials
as in Eq.~(\ref{eq:1VCA}).

To validate our VCA modeled $\mathrm{{Li}}_{x}\mathrm{{TiSe}}_{2}$ system, we
performed calculations for a $3\times3\times1$ supercell Cu-intercalated
$\mathrm{{TiSe}}_{2}$. This corresponds to the Cu doping of $x=0.11$.
The electronic structure was subsequently calculated and unfolded using
the procedure of Refs.\ \cite{Medeiros2014, Medeiros2015}. These unfolded supercell
electronic bands were then compared with the VCA calculations for
$\mathrm{{Li}}_{x}\mathrm{{TiSe}}_{2}$ at $x$ = 0.11.
Fig.~\ref{fig2_DOS} (a) and (b) show the comparison between them.
The good agreement of their dispersions around the Fermi level provides a justification of the VCA.
Also, the electron DOS for the supercell and VCA, as shown in Fig.~\ref{fig2_DOS} (c) and (d), are in good agreement with each other around the Fermi level and also with the previous calculations.\cite{Jishi2008}
The number of doped electrons per unit cell, which is the charge transferred
from the cation to the $\mathrm{{TiSe}}_{2}$ layers, can be calculated
by integrating the DOS for the shifted chemical potential from doping
with respect to the Fermi level of the pristine $\mathrm{{TiSe}}_{2}$.
We found that the same number of electrons ($x=0.11$) per unit cell for both VCA and
supercell calculations as shown by the green shaded areas
in Fig.~\ref{fig2_DOS} (c) and (d).

\begin{figure*}[t]
\includegraphics[scale=0.75]{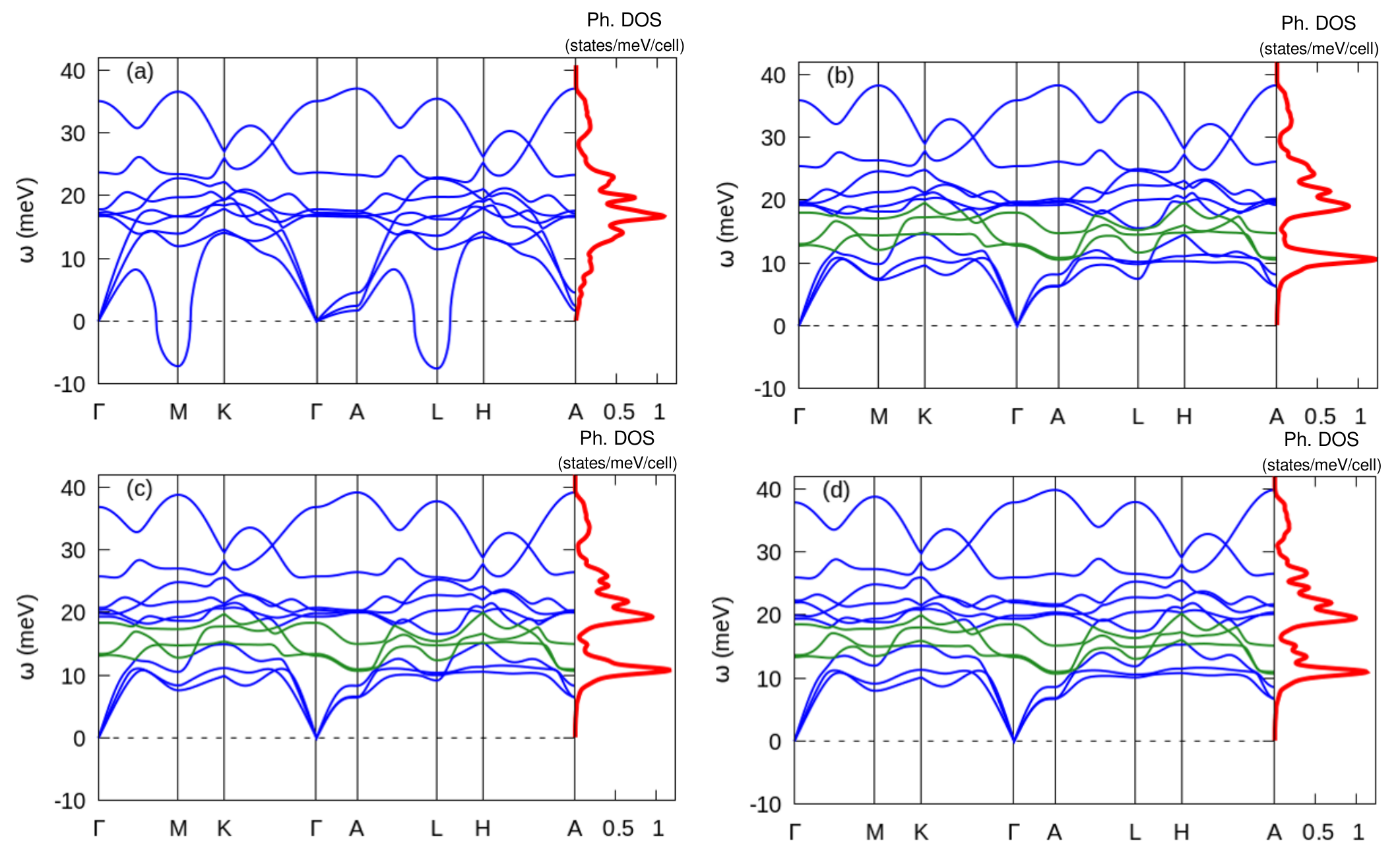}
\caption{ The phonon dispersions and corresponding phonon density of states for $\mathrm{{Cu}}_{x}\mathrm{{TiSe}}_{2}$ at $x$ = 0, 0.05, 0.08, and 0.15 are shown, respectively, in (a), (b), (c), and (d). The dopant induced phonon modes appear in between the acoustic and optical modes and are shown by green colored lines.}
 \label{fig3}
\end{figure*}

For the calculations of the phonon modes of the cation intercalated $\mathrm{{TiSe}}_{2}$,
we replaced the Li ion mass with that of Cu. We then employed DFPT in
the linear response \cite{SBaroni2001} to calculate the dynamical matrices with
$2 \times 2 \times 2$ \textbf{q}-points grid from which to calculate the eigenfrequencies, eigenmodes, and the interatomic force constants.
Then the electron-phonon coupling quantities were calculated on a significantly finer grid using the electron-phonon Wannier (EPW) interpolation scheme.\cite{FGiustino2007, JNoffsinger2010}
The electronic wave functions required for the Wannier-Fourier interpolation were calculated
on a uniform and $\Gamma$ centered \textbf{k}-mesh of size $12 \times 12 \times 12$.
For the maximally localized Wannier orbitals, these six Ti $d_{xy}$, $d_{xz}$, $d_{yz}$ and
Se $p_x$, $p_y$, $p_z$ states were considered.  The electron-phonon matrix elements were first
computed on a coarse $24 \times 24 \times 24$ \textbf{k}-mesh and $2 \times 2 \times 2$ \textbf{q}-mesh.
Then, it was interpolated to a finer $40 \times 40 \times 40$ \textbf{k}-mesh and
a $20 \times 20 \times 20$ \textbf{q}-mesh.

The phonon frequencies $\omega_{\textbf{q}\nu}$ as a function of the phonon wavevector
$\textbf{q}$ for a given phonon mode $\nu$ are calculated by solving the eigenvalue equation,
\begin{equation}
  \text{Det}\vert D^{\alpha \beta}_{I J}(\textbf{q},\nu)- \omega^2_{\textbf{q}\nu}\vert =0.
\end{equation}
$D^{\alpha \beta}_{I J}(\textbf{q},\nu)$ is the dynamical matrix defined as
\begin{equation}
  D^{\alpha \beta}_{I J}(\textbf{q},\nu) = \frac{1}{\sqrt{M_I M_J}}\frac{\partial^2 E_{tot}}{\partial u^{\alpha}_{I}(\textbf{q},\nu)  \partial u^{\beta}_{J}(\textbf{q},\nu)},
\end{equation}
where $E_{tot}$ is the total energy of the system and $u^{\alpha}_{I}$($u^{\beta}_{J}$) is the displacement of atom $I$($J$) in direction $\alpha$($\beta$), and $M_I$($M_J$) denotes the atomic mass.

For a three dimensional material with $N_a$ atoms per unit cell, the number of phonon modes is $3N_a$ and the dynamical matrix becomes $3N_a \times 3N_a$ matrix for $\textbf{q}$ and $\nu$. This requires that the integral of the phonon DOS over frequency be equal to 9 for TiSe$_2$ and $3(3+x)$ for Cu$_x$TiSe$_2$.
This is written in terms of the phonon DOS $F(\omega)$ as
 \ba
\int_0^\infty d\omega F(\omega) = 3(3+x).
\label{sumrule}
 \ea
But, because the Cu intercalation was modelled using the virtual atom, a unit cell has 4 atoms and there exist 12 phonon modes for any nonzero doping. To satisfy the requirement of Eq.\ (\ref{sumrule}), the standard form of the phonon DOS
\begin{equation}
  F(\omega) = \sum_{\nu}\int_{\mathrm{BZ}}\frac{\mathrm{d}\textbf{q}}{\Omega_{\mathrm{BZ}}}\delta(\omega - \omega_{\textbf{q}\nu})
\end{equation}
was extended to
 \ba
  F(\omega) = \sum_{\nu}\int_{\mathrm{BZ}}\frac{\mathrm{d}\textbf{q}}{\Omega_{\mathrm{BZ}}}\delta(\omega - \omega_{\textbf{q}\nu})
 \sum_{j=1}^{N_{ph}} a_{j\nu}^2 \beta_j .
 \ea
$N_{ph}=12$ is the number of phonon modes, $a_{j\nu}$ is the $j$-th component of the eigenvector corresponding to the frequency $\omega_{{\textbf q}\nu}$, and $\beta_j$ equals the concentration $x$ when $j$ refers to the virtual atom and equals 1 otherwise. Then, Eq.\ (\ref{sumrule}) is guaranteed using the normalization condition $ \sum_\nu a_{j\nu}^2 = 1 $.

The electron-phonon matrix element for the scattering of an electron in band $n$ at
wavevector \textbf{k} to a state in band $m$ with wavevector \textbf{k+q} by a
phonon is given by
\begin{equation}
  g_{mn}^\nu (\textbf{k},\textbf{q}) = \left( \frac{\hbar}{2M\omega_{\textbf{q}\nu}} \right)^{1/2}
  \langle m,\textbf{k}+\textbf{q}\vert\delta_{\textbf{q}\nu}V_{SCF}\vert n,\textbf{k}\rangle.
\end{equation}
In this expression, $\vert n,\textbf{k}\rangle$ is the bare electronic Bloch
state, $\omega_{\textbf{q}\nu}$ is the screened phonon frequency, $M$ is the ionic
mass, and $\delta_{\textbf{q}\nu}V_{SCF}$ is the derivative of the self-consistent
potential with respect to a collective ionic displacement corresponding to phonon
wavevector \textbf{q} and mode $\mathbf{\nu}$.

The Eliashberg function, $\alpha^2F(\omega)$, is given by \cite{Allen72}
%
%
 \ba
  \alpha^2F(\omega) = N(\epsilon_F) \sum_{m,n,\nu} \int \int_{\mathrm{BZ}}
 \frac{\mathrm{d}\textbf{k}}{\Omega_{\mathrm{BZ}}} \frac{\mathrm{d}\textbf{q}}{\Omega_{\mathrm{BZ}}}
 \left| g_{mn}^\nu(\textbf{k},\textbf{q}) \right|^2 \nonumber \\
 \times \frac{\delta(\epsilon_{m,\textbf{k}+ \textbf{q}} - \epsilon_F)}{N(\epsilon_F)}
  \frac{\delta(\epsilon_{n,\textbf{k}} - \epsilon_F)}{N(\epsilon_F)}
  \delta(\omega - \omega_{\textbf{q}\nu}) \sum_{j=1}^{12} a_{j\nu}^2 \beta_j,
\label{Eliashbergfun}
 \ea
where $N(\epsilon_F)$ is the density of states at Fermi level per unit cell and per
spin. The dimensionless coupling constant $\lambda$ is given by the integral
\begin{equation}
    \lambda = 2\int_0^\infty \frac{\mathrm{d}\omega}{\omega}\alpha^2F(\omega).
  \label{lambda}
\end{equation}
The SC critical temperature $T_c$ is computed by
the Allen-Dynes-McMillan formula \cite{McMillan68, Allen&Dynes75}:
\begin{equation}
    T_c = \frac{\langle \omega_\text{ln} \rangle}{1.2}\exp\left[
    {\frac{-1.04(1+\lambda)}{\lambda-\mu^*(1+1.062\lambda)}}\right],
  \label{Tc}
\end{equation}
where the logarithmic average of phonon frequency is
\begin{equation}
    \langle \omega_\text{ln} \rangle = \exp\left[\frac{2}{\lambda}\int
    {\frac{\text{d}\omega}{\omega}\alpha^2F(\omega)\text{log} \omega}\right],
  \label{wlog}
\end{equation}
and $\mu^*$ is the Coulomb pseudopotential.
Recall that the two important parameters $\lambda$ and $\langle \omega_{ln} \rangle$ to determine $T_c$ are not independent but are related with each other as given by Eqs.\ (\ref{lambda}) and (\ref{wlog}).

\section{Results and Discussion}


We show in Fig.\ \ref{fig3} the phonon dispersions for pristine $\mathrm{{TiSe}}_{2}$
and doped Li$_x$TiSe$_2$ systems. For the pristine case of Fig.~\ref{fig3} (a) the phonon
dispersion exhibits the imaginary frequencies (shown as negative frequencies) near the $M$ and $L$ points. This indicates the instability toward the $2 \times 2 \times 2$ superstructure mentioned in the Introduction.
Fig.\ \ref{fig3} (b), (c), and (d) show the phonon dispersions for $x=0.05$, 0.08, and 0.11, respectively. The dopant induced phonon modes appear in between the acoustic and optical modes as shown in green lines in the figures.
The doping concentration $x$ was taken into consideration in the $T_c$ calculations via the $\beta_j$ factor in the Eliashberg function of Eq.\ (\ref{Eliashbergfun}).

\begin{figure}[t]
\includegraphics[scale=0.65]{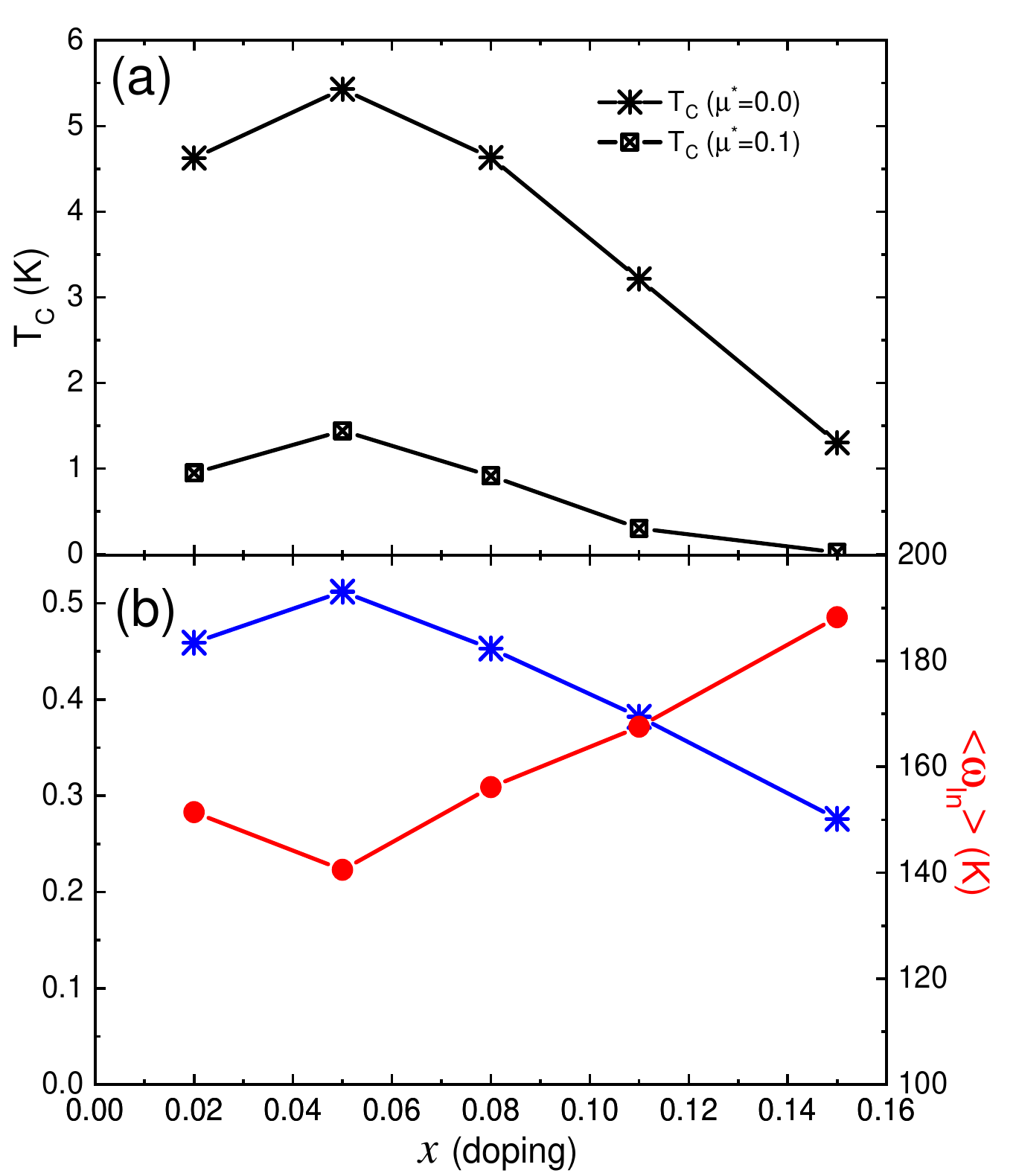}
\caption{
(a) SC critical temperature $T_c$ for $\mu^* = 0.0$ ($\ast$) and $\mu^* = 0.1$ ($\boxtimes$), and (b) logarithmic average of phonon frequency $\langle \omega_\text{ln} \rangle$
(red $\bullet$) and electron-phonon coupling constant $\lambda$ (blue $\ast$), as a variation of doping concentration $x$ in $\mathrm{{Cu}}_{x}\mathrm{{TiSe}}_{2}$.}
 \label{fig4_Tc}
\end{figure}

\begin{figure}[t]
\includegraphics[scale=0.5]{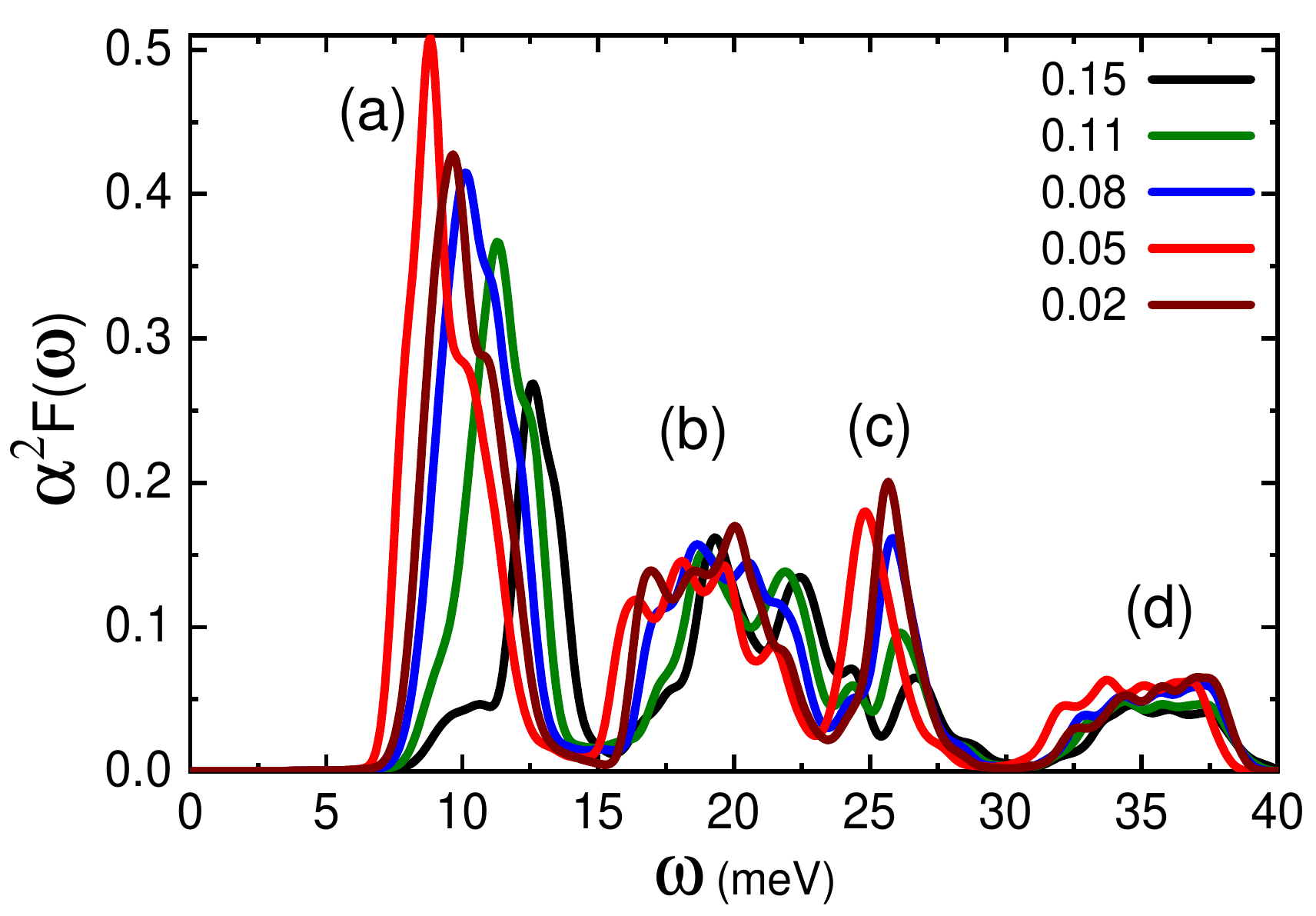}
\caption{
The Eliashberg function $\alpha^2F(\omega)$ calculated using EPW at the dopings indicated in the inset. (a) -- (d) indicate the phonon modes analyzed in Fig.~\ref{fig6}.}
 \label{Eliash}
\end{figure}

In Fig.\ \ref{fig4_Tc}, we show the superconducting $T_c$, the dimensionless electron-phonon coupling constant $\lambda$ and the logarithmically averaged phonon frequency $\langle \omega_\text{ln} \rangle$ as a function of $x$. These were calculated using the Eliashberg function shown in Fig.\ \ref{Eliash} and Eqs.\ (\ref{lambda})--(\ref{wlog}).
Although the calculated doping concentration at the maximum $T_c$ ($x \approx 0.05 \pm 0.01$) is slightly lower than the experimental value ($x \approx 0.077 - 0.08$) due to the overestimated doping effect by the LDA exchange functional,\cite{Jishi2008}
our results reproduce the $T_c$ close to the experimental values and the observed superconducting $T_c$ dome shape as reported in Refs. \cite{EMorosan06,Sawa19}. For the weak coupling superconductors of $\lambda \lesssim 1$, the Allen-Dynes-McMillan formula $T_c$ is very close to the $T_c$ from solutions of the Eliashberg equation.\cite{Allen72}
The obtained $\lambda$ as shown in Fig.\ \ref{fig4_Tc} (b) is about 0.5 or smaller. Maschek {\it et al.} simulated the doping induced CDW suppression and SC in the DFT and DFPT calculations by tuning the smearing parameter $\sigma$, which is often considered as the electronic temperature scale, as the CDW instability was suppressed at $\sigma =150$ meV as $\sigma $ was varied. They reported $\lambda \approx 1$ at the CDW suppression which is about twice larger than the present calculations. This discrepancy probably comes from the different values of $\sigma$ and different treatment of the doping. We took $\sigma = 10$ meV. In the current calculations we explicitly included the dopants in the calculations which suppress the CDW instability and there was no need to tune the smearing parameter to large values.

The maximum $T_c$ as a function of $x$ is pinned to the minimum phonon frequency as has been observed and discussed in many classes of materials.\cite{Testardi_RMP1975,Moussa_PRB2006}
The reduced phonon frequency enhances $\lambda$ due to the $\omega$ factor in the denominator of Eq.\ (\ref{lambda}). Recall that a reduced frequency decreases the $T_c$ from the viewpoint of BCS $T_c$ formula. But the enhanced $\lambda$ overcompensates the frequency decrease to produce a net increase of $T_c$. From a more general viewpoint which is valid for the strong coupling $T_c$ as well, we have
 \ba
T_c \sim \sqrt{\omega_0^2 -\omega_{ph}^2},
 \ea
where $\omega_0$ and $\omega_{ph}$ are, respectively, the bare and renormalized phonon frequency.\cite{Moussa_PRB2006} $T_c$ is maximum when $\omega_{ph}$ is minimum.
As doping is increased or decreased away from the optimal concentration, the phonon frequency hardens back and $\lambda$ is decreased to yield a reduced $T_c$. This results in a dome shape of $T_c$ as a function of $x$ in Cu$_x$TiSe$_2$. When $\lambda$ becomes comparable with the Coulomb pseudopotential $\mu^*$, superconductivity is completely suppressed which is around $x \approx 0.15$ for $\mu^* =0.1$ without the doping induced scattering enhancement as $x$ increases.\cite{Zhao_PRL2007}

This discussion can be seen more explicitly in
Fig.\ \ref{Eliash} which shows the electron-phonon coupling Eliashberg function $\alpha^2 F(\omega)$ for the doping concentration $x=0.02$, 0.05, 0.08, 0.11, and 0.15.
The $\alpha^2 F(\omega)$ displays four groups of modes as labelled as (a), (b), (c), and (d) in the figure. The most significant variation of $\alpha^2 F(\omega)$ as a function of $x$ is the peak position and strength of mode (a). It exhibits the lowest frequency and largest strength at $x=0.05$ to give the highest $T_c$.

\begin{figure}[b]
\includegraphics[scale=0.55]{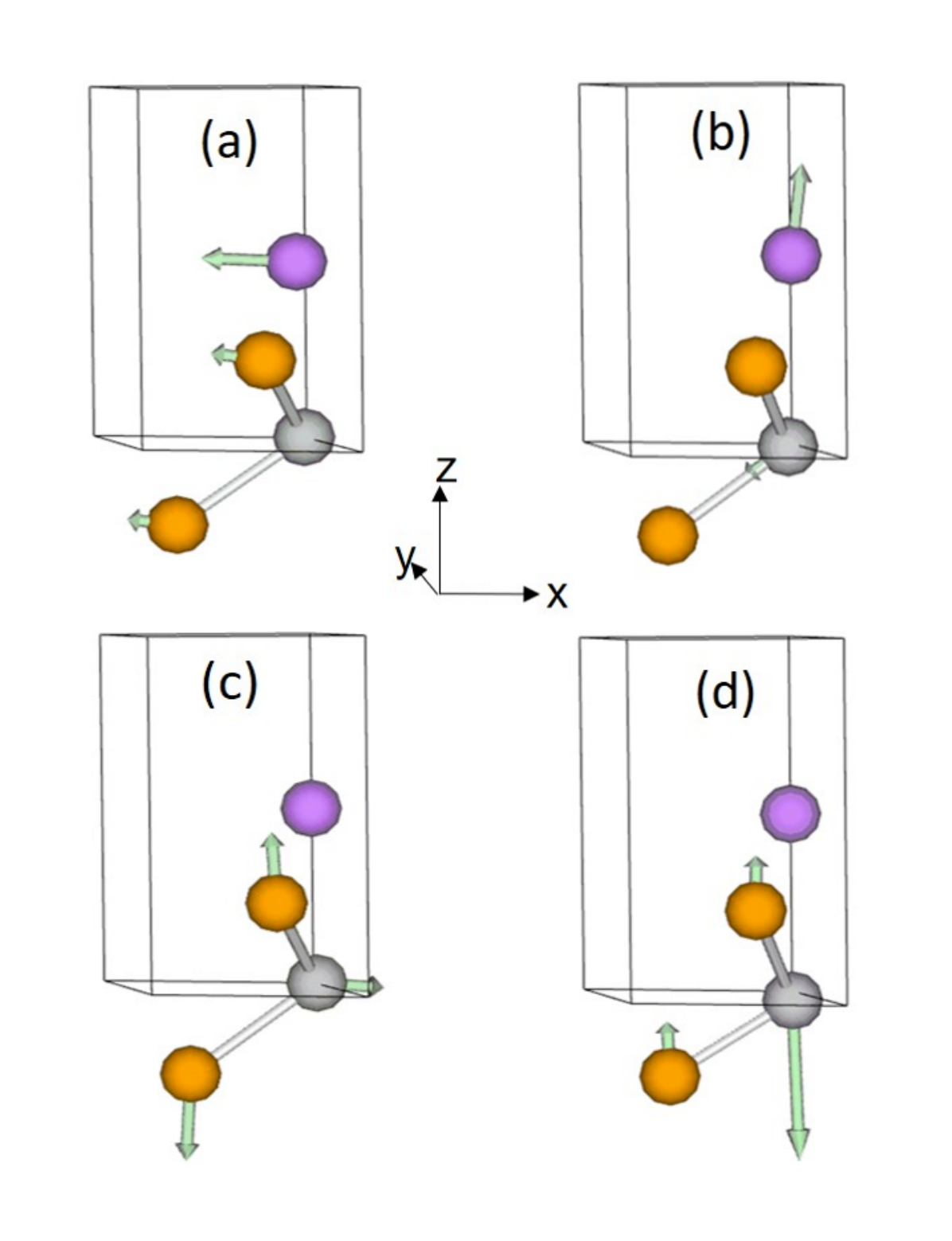}
\caption{\label{fig6}
The analysis of the phonon modes contributing to superconductivity. The labels (a) -- (d) correspond to the modes of Fig.~\ref{Eliash}. The grey, yellow, and purple balls represent the Ti, Se, and virtual Li atoms, respectively. The green arrows represent the direction of the vibration motion of atoms.}
\end{figure}

The character of the modes is shown in Fig.\ \ref{fig6}.
The mode Fig.~\ref{fig6} (a), also labelled as (a) in Fig.\ \ref{Eliash}, is characterized by the in-plane dopant ion
and Se ion vibrations. This is the combination of the longitudinal acoustic and
the first dopant optical mode.\cite{Takaoka_JPSJ1980} It
contributes about $2/3$ of the total $\lambda$ and is the dominant phonon mode contributing to the superconductivity. The (b) phonon mode is a combination of the out-of-plane dopant optical mode and Ti transverse optical mode.\cite{Jaswal_PRB1979}
These modes of (a) and (b) involving the dopant ions would not have been captured in the calculations, for example, using the rigid shift of the Fermi energy.
The phonon mode (c) is characterized by the combination of the out-of-plane Se and the in-plane Ti displacements.
This mode dominantly comes from the longitudinal optical phonon mode.\cite{Takaoka_JPSJ1980,WAKABAYASHI1978}
Finally, the phonon mode (d) in the high energy region is the out-of-plane longitudinal optical mode
of the out-of-phase Ti and Se ions.\cite{Tornatzky_PRB2019}

We have presented that the doping induced SC in Cu$_x$TiSe$_2$ may be understood in terms of the phonon mediated pairing. The maximal $T_c$ was pinned to the CDW QCP at which the phonon frequency is minimum and $\lambda$ is maximum.
On the other hand, the pressure induced SC seems at odds with this picture, although there are reports that this can also be understood within the phonon mediated pairing.\cite{Calandra2011} The CDW is completely suppressed, that is, $T^{CDW} = 0$ if the normal state gap $E_g$ is larger than the exciton binding energy in the excitonic insulator picture. The exciton condensation driven CDW is also suppressed by a Lifshitz transition (that is, $E_g = 0$) reported by Bok {\it et al.} from their material specific calculations.\cite{Bok_PRB2021} Accepting that $E_g $ is positive (semiconducting) at the ambient pressure for TiSe$_2$, the pressure induced QCP at $P_c \approx 5.1 $ GPa suggests that $E_g =0$ at $P = P_c$ and $E_g > 0$ for $P<P_c$.

The superconductivity for $2 \lesssim P \lesssim 4$ GPa then emerges out of a semiconducting normal state. This can be the $s_\pm$ SC state suggested by Ganesh {\it et al.} if samples are accidentally doped, for example, by crystal imperfections.\cite{Ganesh_PRL2014}
On the other hand, Maschek {\it et al.} argued that the hybridization between the phonon and exciton modes pushes the critical point of the hybrid mode up but $T_c$ maximum is below the QCP. This, however, does not seem to be supported by the recent work by Lee {\it et al.}\cite{Lee_PRR2021}
They reported that a SC dome appears centered at the tunable QCP by both the Cu concentration and pressure.
Alternatively, if the incommensurate CDW intervenes before the CDW is completely suppressed as reported by Kogar {\it et al.} for Cu intercalation\cite{Kogar_PRL2017} and by Joe {\it et al.} for the pressure,\cite{YIJoe14} the incommensurateness is accommodated by forming the domains walls. Then SC may emerge out of the domain wall metallic state via phonon mediated pairing for both doping and pressure induced cases. It seems highly desirable that the experimental phase diagram in the doping and pressure parameter space is more accurately delineated with regard to the commensurate and incommensurate CDW phases.

\section{Summary and concluding remarks}

In this paper, we demonstrated that the superconducting $T_c$ as a function of the Cu intercalation concentration $x$ for Cu$_x$TiSe$_2$ is entirely determined by the electron-phonon interaction without a necessity to invoke the exciton mechanism.
We employed the {\it ab initio} density functional theory and density functional perturbation theory to calculate the electron-phonon coupling Eliashberg function $\alpha^2 F(\omega)$ from which to calculate the $T_c$ using the Allen-Dynes-McMillan formula.
The maximum $T_c$ was found to be pinned to the minimum phonon frequency as has been observed and discussed in many classes of materials.\cite{Testardi_RMP1975,Moussa_PRB2006}

The phonon mechanism of superconductivity for Cu$_x$TiSe$_2$ is not at odds with the excitonic insulator picture of the CDW state. Note that the excitonic and lattice instabilities should appear simultaneously as reported by Kogar {\it et al.} because they have the same spatial symmetry.\cite{Kogar_Science2017,Bok_PRB2021} Within this picture, the excitonic instability drives the CDW formation of the $2\times 2\times 2$ periodic lattice distortion. Then, the CDW must be accompanied by the soft phonon mode of the wavevector $L$ at the CDW quantum critical point. The critical temperature $T_c$ becomes maximum at the minimum phonon frequency within the Eliashberg formalism as we obtained here.\cite{Moussa_PRB2006} The superconductivity is phonon mediated conventional pairing and the proximity of the maximal superconductivity and the CDW quantum critical point is not coincidental.

\begin{acknowledgments}

We acknowledge valuable discussions with Hyoung Joon Choi with regard to the virtual crystal approximation. This work was supported by the National Research Foundation of Korea under NRF-2020R1I1A1A01054852 (THP) and NRF-2021R1F1A1063697 (HYC).

\end{acknowledgments}

\bibliographystyle{unsrt}

\bibliography{VCATiSe2}

\end{document}